\documentclass[journal]{IEEEtran}

\IEEEoverridecommandlockouts                              

\usepackage{graphics} 
\usepackage{epsfig} 
\usepackage{mathptmx} 
\usepackage{times} 
\usepackage{amsmath} 
\usepackage{amssymb}  
\usepackage{bm}
\usepackage{subfigure}
\usepackage{url}
\usepackage{color,soul}
\usepackage{ulem}
\usepackage{color}
\usepackage{xcolor}
\usepackage{paracol}
\usepackage[colorlinks,
linkcolor=blue,
anchorcolor=blue,
citecolor=blue,
]{hyperref}
\definecolor{keywordcolor}{rgb}{0.8,0.1,0.5}
\usepackage{listings}
\lstdefinelanguage{JavaScript}{
	morekeywords=[1]{break, continue, delete, else, for, function, if, in,
		new, return, this, typeof, var, void, while, with, require},
	morekeywords=[2]{false, null, true, boolean, number, undefined,
		Array, Boolean, Date, Math, Number, String, Object},
	morekeywords=[3]{eval, parseInt, parseFloat, escape, unescape},
	sensitive,
	morecomment=[s]{/*}{*/},
	morecomment=[l]//,
	morecomment=[s]{/**}{*/}, 
	morestring=[b]',
	morestring=[b]"
}[keywords, comments, strings]

\bstctlcite{IEEEexample:BSTcontrol}

\title{Svar: A Tiny C++ Header Brings Unified Interface for Multiple programming Languages}

\author{Yong Zhao , Pengcheng Zhao, Shibiao Xu, Lin Chen, Pengcheng Han, Shuhui Bu, Hongkai Jiang
\thanks{Yong Zhao, Pengcheng Zhao, Lin Chen, Pengcheng Han, Shuhui Bu and Hongkai Jiang are with Northwestern Polytechnical University, 710072 Xi'an, China. Shibiao Xu is with School of Artificial Intelligence, Beijing University of Posts and Telecommunications, and Key Lab of Universal Wireless Communications, Ministry of Education. Shibiao Xu is the corresponding author (shibiaoxu@bupt.edu.cn).}%
}

\begin{document}

\maketitle
\thispagestyle{empty}
\pagestyle{empty}

\begin{abstract}
There are numerous types of programming languages developed in the last decades, and most of them provide interface to call C++ or C for high efficiency implementation.
The motivation of Svar is to design an efficient, light-weighted and general middle-ware for multiple languages, meanwhile, brings the dynamism features from script language to C++ in a straightforward way.
Firstly, a Svar class with JSON like data structure is designed to hold everything exists in C++, including basic values, functions or user defined classes and objects.
Secondly, arguments are auto cast to and from Svar efficiently with compile time pointers, references and shared\_ptr detection.
Thirdly, classes and functions are binded with string names to support reflection, this means all functions and classes in a shared library can be exported to a Svar object, which also calls a Svar module.
The Svar modules can be accessed by different languages and this paper demonstrates how to import and use a Svar module in Python and Node.js.
Moreover, the Svar modules or even a python module can also be imported by C++ at runtime, which makes C++ more easier to compile and use since headers are not required anymore.
We compare the performance of Svar with two state-of-the-art binding tool for Python and Node.js, and the result demonstrates that Svar is efficient, elegant and general.
The core of this project is one single tiny modern C++ header with less than 5000 lines code without extra dependency.
To help developers using Svar, all the source codes related are public available on github \url{http://github.com/zdzhaoyong/Svar}, including documentations and benchmarks.

\end{abstract}

\section{Introduction} \label{sec_intro}

For this world built on coding, most languages use C/C++ as their core implementation or provide interface to call C/C++ libraries for higher efficiency.
New programming languages are still coming and they all has their particular superiority \cite{cass20182017, harper2016practical}.
It would be great if we can import libraries across languages, and a C++ based middle-ware is the best option to maintain efficiency and maximize the community. 

One challenge is the type translation between weakly and strongly typed languages.
Being a natively compiled language, C++ can not generate other C++ code at runtime.
Reflection in a C++ context can only rely on external scripting languages integrated into the C++ project.
However, given the capabilities of dynamic objects to expose themselves, it is also very easy to expose them to a scripting language.
Therefore, dynamism is the key role to provide full reflection for strongly typed languages, where variable, functions and classes are all represented in an uniform way.
And dynamism is not only beneficial for scripting language integration, but also brings advantages like more abstraction, faster compiling and less ugly hacks.

Many binding tools like SWIG \cite{beazley1996swig}, pybind11 \cite{jakob2017pybind11}, nbind \cite{charto2017nbind} are developed to help the interaction between different languages and C++.
However, existing binding tools need users to develop a particular wrapper for each language and brings extra dependency.
A unified interface would help developers to write less wrapper code and useful for library developing and releasing.


%



\begin{figure}[tb]
	\centering
	\includegraphics[width=0.48\textwidth]{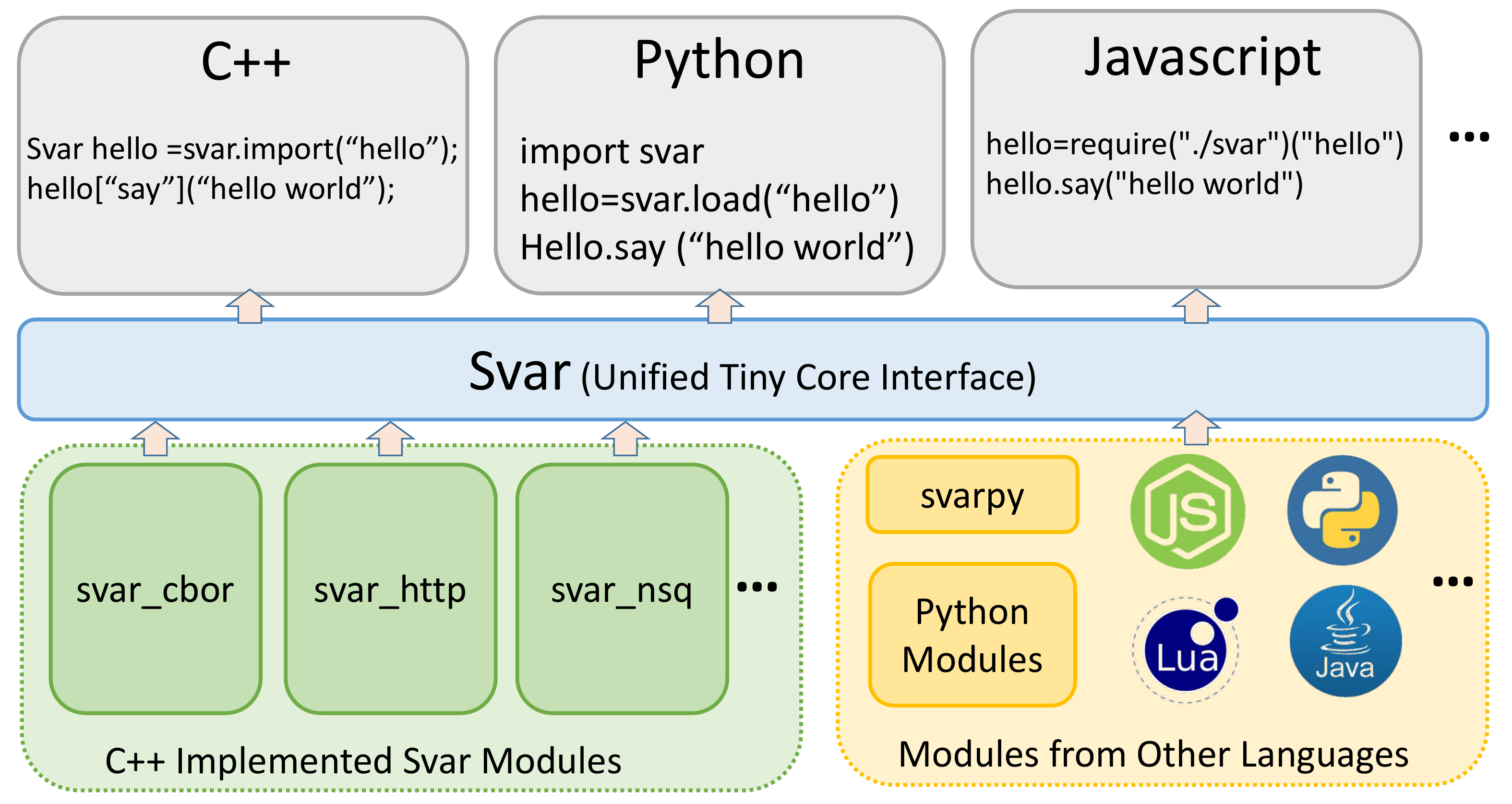}
	\caption{Svar is an efficient, light-weighted and general middle-ware for multiple languages.
		Firstly, for C++ programmers, it provides more dynamism for new design pattern and a general form to import shared libraries dynamically, which makes C++ programming, compiling and releasing easier.
		Secondly, it unifies different C++ binding tools in an elegant style where multiple languages are supported at the same time and further brings JSON supporting.
		Thirdly, Svar is designed to be a high performance bridge for module importing among different languages.
	}
	\label{fig:cover}
\end{figure}

As illustrated in Fig. \ref{fig:cover}, the motivation of Svar is to design an efficient, light-weighted and general middle-ware for multiple languages, meanwhile, brings the dynamism features from script language to C++ in a straightforward way.

In brief, the main contributions are concluded as follows:
\begin{enumerate}	
	\item \textbf{A tiny header file that makes C++11 library accessible from other programming languages.} Due to the high efficiency of C++/C, other languages often use it for low level functional implementations. They often need a wrapper for helping calling C++ through the language interface protocol. Svar is designed to be a high efficient middle-ware between languages, and the porting of a language can be done by just easily porting the single Svar class. Our experiments shows that Svar maintains high efficiency and is even faster than traditional binding tools. The porting of Svar is reusable and the C++ implementation will not brings extra dependency, which also means the C++ library can be used in different languages at the same time without modification.
	
	\item \textbf{Enables \textit{import} programming style instead of \textit{include} in C++. } Shared libraries are imported inside without header dependency, makes C++ programming and compiling easier. As too many headers slow down the compilation speed or even brings compile errors, C++ beginners or even experienced programmers often take long time fixing compiling problems, which makes it difficult or even impossible to use a lot third-party libraries in C++. Svar allows to import the libraries without dependencies or headers, which means modules are decoupled and compile a large project with tens of dependencies has the same complexity as a "Hello World" project.
	
	\item \textbf{A generic holder for everything.} The standard template library (STL) in C++ only allows to hold specific type which is already known, when developers want to hold an object with dynamic type, they don't have a proper solution. Although the C++17 standard brings std::any and std::variant for holding different values, they are not able to be used without original type declare. Svar brings an easy and safe way to hold everything including functions and classes, while maintains usability without declaring. The contents can further be organized in a JSON style structure, which popular used in script languages. This dynamic feature brings more different design patterns which makes programming easier.
	
	\item \textbf{A high performance bridge among different languages.} As every language has its metrics in particular areas, some functions are often implemented by different languages. 
	It would be great if we can import libraries across languages.
	A C++ middle-ware is the best option to maintain efficiency and maximize the community.
	Svar not only helps calling C++ in other languages, but also allows to import libraries implemented by other languages, which forms a bridge among programming languages.
	
\end{enumerate}

\section{Related Works} \label{sec_related}

\textbf{Language binding tools.} Programming languages generally provide interface to call some other native languages like C, and lot of binding tools are developed to simplify wrapping work as the low level interface requires a considerable amount of effort and expertise.
SWIG \cite{beazley1996swig} is a custom C++ parser to help developers generating wrapper for languages like Python, Perl and Java.
Rather than a custom language, Boost.Python \cite{abrahams2008boost} uses C++ wrapper files instead and specifies the classes and (member) functions explicitly.
This makes it easier for C++ developers to read, write and understand than Swig, and allows wrappers to be made more Pythonic.
Pybind11 \cite{jakob2017pybind11} further improved Boost.Python in a more elegant style with less code by using modern C++11 standard. 
Other binding tools are also implemented for languages like JavaScript \cite{charto2017nbind} and Lua \cite{lua-intf}.
However, all those binding tools brings extra dependency of languages and developers need proficiency in both source and target languages.
Svar is designed to decouple those dependencies, where developers are able to expose everything without considering target language and call libraries across languages seamlessly.

\textbf{Source code translators and compilers.} Several studies have investigated the possibility to translate programming languages for better compatibility or efficiency. 
Cython \cite{behnelcython} tries to extend Python and compile scripting source code to C, makes writing tiny accelerated C extensions for Python easier. 
Instead of define static types explicitly, Shed \cite{shedskin} uses type inference techniques to determine the implicit types used in a Python program and translate typed Python programs into optimized C++.
Pythran \cite{guelton2015pythran} further enables static optimization of scientific Python programs by using explicit thread-level parallelism through OpenMP annotations, false variable polymorphism pruning, and automatic vector instruction generation such as AVX or SSE.
With developing of deep learning, supervised \cite{chen2018tree} and unsupervised \cite{lachaux2020unsupervised} methods are also proposed to translate programming languages.
Most programming language translators are experimental with too may limitations for practical usage now due to difficulties handling preprocessing, classes, templates, and all the idiosyncrasies and complexities of languages.
Assembly is more verbose, lower-level and simpler to work on.
Emscripten \cite{zakai2011emscripten} presents a low level compiler base on LLVM (Low Level Virtual Machine) which can compile C and C++ into JavaScript, and opens up new opportunities for running code on the web.
	
\textbf{Reflection, dynamism and module design in C++.} 
As most languages (Java, Python) support reflection as part of their standard specifications, C++ only support very limited reflection features though RTTI (Runtime Type Information) \cite{skochinsky2012compiler}.
Some early work \cite{jautzy1997metalevel,chiba1995metaobject} tried to enhance C++ with reflection capabilities using MOP (Metaobject Protocol). However, these approaches are either intrusive or are not fully compliant with the C++ standard. 
More reflection work \cite{rttr,devadithya2007c++} which fully compliant with the standard C++ specification are implemented and likely the recent standards of C++ (i.e., C++2x) will have support for compile time reflection.
As dynamism is the key feature to solve runtime reflection, Svar implemented a dependency free single header reflection with dynamism features which is more easier to use.
Moreover, the reflection also brings an unified interface for multiple other programming languages and C++ itself by a modern import style.
 


\begin{figure*}[tb]
	\centering
	\includegraphics[width=0.98\textwidth]{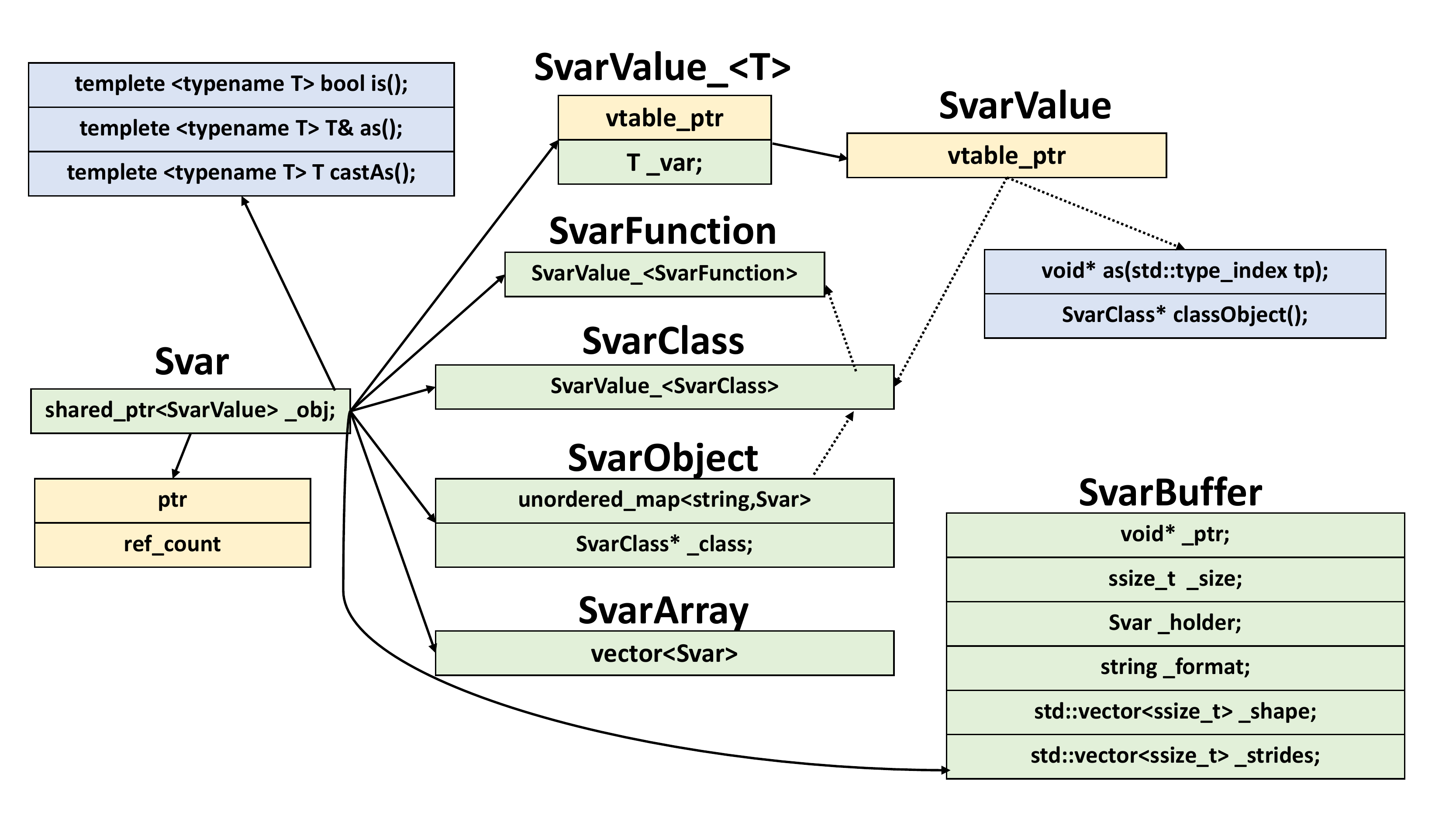}
	\caption{The relationship between basic data structures of Svar, which is simple, efficient and safe.		 
		As \textit{shared\_ptr} is used to maintain the life-cycle of objects, developers do not need to care the memory releasing things. 
		And compare to \textit{shared\_ptr\textless T\textgreater}, the extra cost is only a virtual table pointer, which means holding objects to Svar instead of \textit{shared\_ptr} brings powerful dynamism but negligible memory usage.
	}
	\label{fig:data_structure}
\end{figure*}

\section{Implementation} \label{sec_impl}  
	
Svar is designed to represent everything in C++ and a dynamic type object holder is firstly designed for both performance and usability.
We extend the JSON data structure to organize any possible objects including user-defined functions, classes and instances.
For function calling, arguments are auto cast to and from Svar efficiently with compile time pointers, references and shared\_ptr detection.
Classes and functions are binded with string names to support reflection, this means all functions and classes in a shared library can be exported to a Svar object, which also calls a Svar module.
The Svar modules can be accessed by different languages and this paper demonstrates how to import and use a Svar module in C++, Python and Node.js.
For any other languages, developers only need to simply implement an interface to the single class Svar which can be reused by all modules, instead of repeat binding each object.
Moreover, the Svar modules or even a python module can also be imported by C++ at runtime, which makes C++ more easier to compile and use since headers are not required anymore.
This section will describe how Svar works for C++, as binders of other languages also use C++ interfaces and Svar provides better accessing.

\subsection{Basic Data Structure} 

The data structure of Svar is designed to bring dynamism and auto GC (Garbage Collection) \footnote{\url{https://en.wikipedia.org/wiki/Garbage_collection_(computer_science)}}, which are both core features in scripting languages.
Although C++17 standard brings \textit{std::any} and \textit{std::variant}, they are only a safer naked pointer and no operators are supported.
As C++ only provides very limited dynamism though virtual functions and RTTI, Svar use them to extend dynamism where more informations and operators available.
In the design of most programming languages like Java, Python and JavaScript, the GC is used to automatically manage memory, which frees the programmer from manually dealing with memory deallocation. As a result, certain categories of bugs are eliminated or substantially reduced.
The RAII (Resource Acquisition Is Initialization) technique in C++ binds the life cycle of resources that must be acquired before use and releases them in their destructors.
And smart pointers like \textit{unique\_ptr} and \textit{shared\_ptr} are provided in modern C++ to manage dynamically-allocated memory.
As Svar is a dynamic type holder where contents are generally dynamically-allocated, \textit{shared\_ptr} is used to provide garbage collection.
The basic data structure of Svar is illustrated in Fig. \ref{fig:data_structure}, where shows Svar is designed to be simple, efficient and safe.
As \textit{shared\_ptr} is used to maintain the life-cycle of objects, developers do not need to care the memory releasing things. 
And compare to \textit{shared\_ptr\textless T\textgreater}, the extra cost is only a virtual table pointer, which means holding objects to Svar instead of \textit{shared\_ptr} brings powerful dynamism but negligible memory usage.

\subsection{JSON Structure Dynamic Holder} 

Svar is a dynamic holder for everything, where anything can be hold by Svar.
Moreover, a JSON style is used to organize objects, which brings powerful abilities to easily represent and access any data structure.
Some examples are given in the following C++ code:

\begin{lstlisting}[caption={Svar is a JSON structure dynamic holder for everything.},label=code_demo]
struct A{int a;}
Svar null=nullptr;
Svar f=false;
Svar i=1;
Svar d=2.1;
Svar s="hello world";
Svar a=A({1}); // user defined class instance
Svar v={1,2,3};
Svar m={{"b",false},{"s","hello"},{"n",nullptr},{"u",Svar()}};
m["pi"]=3.14159;
std::cout<<v[2]<<","<<m["pi"]<<std::endl;

double dv[9]; // 3x3 double matrix
Svar b=SvarBuffer(dv,{3,3}).clone(); 
if(s.is<std::string>()) // check type
 std::string str = s.as<std::string>();

try{
 double db=i.as<double>(); // throw
}
catch(std::exception& e){
 double db=i.castAs<double>(); // ok
}

for(auto it:v) // array iterator visit
 std::cout<<it<<std::endl;

for(std::pair<std::string,Svar> it:m)
 std::cout<<it.first<<":"<<it.second<<"\n";

std::string json=m.dump_json();
sv::Svar obj=Svar::parse_json(json);
\end{lstlisting}

The above code demonstrates that JSON is supported as a subset of Svar, and user-defined classes can also be hold by Svar like scripting languages.
Type information can be checked, while operators and iterators are provided to access object easily. 
As pointer and \textit{shared\_ptr} instead of value are also popular used in C++, they are supported to be hold and treated just like their original meta type.

\begin{lstlisting}[caption={Svar holds value, pointer, \textit{unique\_ptr} and \textit{shared\_ptr} with auto casting support.},label=code_demo]
struct A{int a,b,c;}; // define a struct

A a={1,2,3};
Svar avar=a; // value copy
Svar avar=A({1,2,3}); // copy free rvalue
Svar aptrvar=&a;  // pointer assign
Svar uptrvar=std::make_unique<A>({2,3,4});
Svar sptrvar=std::make_shared<A>({2,3,4});

A& ref0=avar.as<A>();
A& ref1=aptrvar.as<A>();
A& ref2=uptrvar.as<A>();
A& ref3=sptrvar.as<A>();

A* aptr=avar.castAs<A*>();
aptr   =aptrvar.as<A*>();
aptr   =uptrvar.castAs<A*>();
aptr   =sptrvar.castAs<A*>();

auto  auptr=uptrvar.as<std::unique_ptr<A>>();
auto  asptr=sptrvar.as<std::shared_ptr<A>>();
\end{lstlisting}

\subsection{Function Holding and Auto Arguments Casting} 

Function plays the most important role in all programming languages, as some shell scripts only use string values and functions to work.
Svar is not only designed to represent and use functions in C++, but also functions objects in other programming languages.
Firstly, all forms of C++ functions including plain C functions, member functions, static member functions and functor objects (including lambda expressions) are supported. 
Secondly, we further brings support of keyword arguments and meta attributes to C++ as they are popular used in scripting languages.
Thirdly, overload and auto arguments type casting are supported to correctly call the functions as expected.
The function signatures are statically auto determined at compile time, and checked at runtime when the function being called.
Some very simple example usages are listed as below:

\begin{lstlisting}[caption={Svar holds different functions and auto cast arguments. Keyword arguments and extra documentations are further supported.},label=code_demo]
int add(int a,int b){
 return a+b;
}

struct A{
 int a;
 void print(){std::cout<<a<<std::endl;}
 static A create(int v){return A({1});}
}

void demo_svar_functions(){
 Svar f = add; // plain C function
 assert(f(1,2).as<int>()==3);
 
 Svar lambda = [](const int& a, int* b){
  return a+(*b);
 }; // auto type casting
 assert(lambda(1,2)==3);
  
 Svar static_func=&A:create;
 Svar a = static_func(1.2); // int -> double
 
 Svar mem_func=&A::print; 
 mem_func(a); // auto cast A -> A*
 
 Svar kw_f(add,"a"_a,"b"_a=0,"add two int");
 assert(kw_f(1,2)==3); // call with arguments
 assert(kw_f(3)==3); // b default is 0
 assert(kw_f("b"_a=1,"a"_a=2)==3); // kwargs
 
 kw_f.overload([](int a,int b,int c){
  return a+b+c;
 }); // overload is supported
 assert(kw_f(1,2,3)==6);
}
\end{lstlisting}

Comparing to the original C++ functions, the Svar form has some advantages and can be used in some particular circumstances.
Firstly, the unified form can represent different functions, which is more suitable for functions with undefined arguments. 
Secondly, the Svar representation maturely supports reflection, and a key-value dictionary can further represents a function package where function signatures and documentations can be maintained in the meta.
Thirdly, keyword arguments are supported for more easier argument usage, and the general form can be easily porting to other languages.

\subsection{Class Reflection and Usage} 

Being able to access class information at runtime increases the flexibility in programming, since it allows developing generic applications like object instantiation, serialization and persistence.
The factory method pattern is widely used to polymorphically instantiate objects whose concrete types are not known at compile-time. However, this pattern nature burdens programmers by requiring to implement such methods for new and already existing classes. Also, the factory method pattern requires all the candidate classes to have a common base class. An approach using reflection would not have such restrictions and will not be intrusive.
Serializing an object state to a byte sequence and de-serializing it back is a common approach to communicate between loosely coupled applications. With reflection, the sender is able to serialize an object's state and sent it to the receiver so that the receiver is able to dynamically create the objects using their descriptions and populate them with the corresponding state information.
Serialization also enables a generic object persistence framework, since it is just a matter of saving the serialized content to a persistent media.
Moreover, a reflection-based binding would minimize the effort required for language binding. In this case, bindings are required only for the reflection related routines, which can easily be automated.
Svar implemented a MOP based class reflection which is general and easy to use:

\begin{lstlisting}[caption={Reflection of C++ classes using Svar.},label=code_demo]
class Person{
public:
Person(std::string name): _name(name){}

static Person create(std::string name){
 return Person(name);
}

virtual std::string info() const{
 return Svar({{"name",_name}}).dump_json();
}
std::string _name;
};

class Student: public Person{
public:
Student(std::string name,std::string school)
 : Person(name),_school(school){}
 
virtual std::string intro()const{
 return Svar({{"name",_name},{"school",_school}}).dump_json();
}
std::string _school;
};

void demo_class_reflection()
{
 // define the class to Svar
 Class<Person>("Person","The base class")
  .construct<std::string>()
  .def("info",&Person::info)
  .def_static("create",&Person::create)
  .def_readonly("name",&Person::_name)
  .def("rename",[](Person& self,string name){
   self._name=name;
  }); // labmda expression

 Class<Student>("Student")
  .construct<std::string,std::string>()
  .inherit<Person>()
  .def_readwrite("school",&Student::_school);

 // use the class with Svar
 Svar Person =svar["Person"]; // class object
 Svar Student=svar["Student"];

 Svar v_mom   = Person("mom");
 Svar v_me    = Student("me","nwpu");
 Person&  mom = v_mom.as<Person>();
 Student& me  = v_me.as<Student>();
 assert(v_mom.call("info") == mom.info());
 assert(v_me.call("info") == me.info());
 assert(v_mom.get("name") == mom.name);
 v_mom.call("rename","mom");
 v_me.set("school","nwpu");
}
\end{lstlisting}

The above example demonstrates that classes and instances are all represented by Svar, and instances constructed with Svar can be further transformed back to static type C++. Except for wrapping existing C++ classes, Svar also supports to define a class dynamically, which enables to represent class objects imported from other languages. Below is a simple sample to define and use a dynamic class:

\begin{lstlisting}[caption={Define and use a class dynamically.},label=code_demo]
void demo_dynamic_class()
{
 Svar Person=SvarClass("Person")
  .def("__init__",[](Svar self,string name){
   self["name"]=name;
  }) // constructor
  .def("info",[](Svar self){
   return self.dump_json();
  }); // member function

 Svar v_mom   = Person("mom");
 std::cout<<v_mom.call("info")<<"\n";
}
\end{lstlisting}

\subsection{Class Member Operators} 

Class operators is supported by most programming languages as it explicitly simplified code writing especially for scientific computing. 
Svar use the Python style operator name definitions where arithmetic, comparison and bitwise operators are defined by a serials member functions with particular name signature.
Here is a simple example to demonstrate how Svar supports operators in C++:

\begin{lstlisting}[caption={Example usage of class operators in Svar. },label=code_demo]
#include <Svar/Svar.h>
#include <complex>
using namespace sv;
using namespace std;
typedef std::complex<double> Complex;

int main(){
 Class<Complex>("Complex")
  .construct<double,double>()
  .def("__add__",[](Complex& self,Complex& r){return self+r;}) // +
  .def("__eq__",[](Complex& self,Complex& r){return self==r;}); // ==

 Svar C=SvarClass::instance<Complex>();
 Svar a=C(1,2);
 Svar b=C(2,3);
 assert((a+b)==Complex(3,5));
 return 0;
}
\end{lstlisting}

\subsection{Svar Module Exporting and Importing} 

Instead of individual variable, function or class objects, a library is combination of them, and a Svar module organize them with a key-value dictionary.
This means the whole library is reflection supported and the module contents can be accessed with names, which naturally forms an unified interface for C++.
Once shared library elements are exposed to the dictionary, users can view the API documentation and use the module dynamically without headers, which simplified SDK distribution and compilation. 
Moreover, the Svar module can further be directly imported by other languages, while module testing and using with scripting languages effectively reduce the burden of development. Here below is a simple example shows how to export C++ library to Svar module:

\begin{lstlisting}[caption={Sample to export C++ library to Svar module. The source code should be compiled to a shared library.},label=code_demo]
#include <Svar.h>

using namespace sv;

void say(std::string v){
 std::cerr<<v<<std::endl;
}

class Person{
public:
Person(std::string name): _name(name){}

virtual std::string info() const{
 return Svar({{"name",_name}}).dump_json();
}
std::string _name;
};

REGISTER_SVAR_MODULE(hello){
 svar["__doc__"]="Sample C++ based module";
 svar["say"] = say; // export function
 svar["version"]=1; // export variable
 
 Class<Person>("Person","The base class")
  .construct<std::string>()
  .def("info",&Person::info);
}
EXPORT_SVAR_INSTANCE
\end{lstlisting}

The above tiny demo shows that exporting variable, function or class to Svar is simple and easy to understand.
Developers just need to compile the source code like ordinary shared libraries, and it can be easily used in different languages, while the exporting does not rely on the low level data structure or dependencies of those programming languages.
This not only simplified the wrapper binding works for scripting languages using, but also removed extra language dependency and repeated wrapping, so that all languages can shares the same implementation for designing, testing and releasing.
Moreover, for C++ programmers, the shared library can either be linked at compile time with headers, but also be used without interface headers and be imported at runtime like scripting languages.
API documentations about the interface can be illustrate using the built-in command tool or directly print the Svar object contents.
Here is a demo shows how to import and use the above exported shared library dynamically:

\begin{lstlisting}[caption={Demo C++ code import and use shared library Svar module.},label=code_demo]
#include <Svar.h>

int main(){
  auto hello = svar.import("hello");
  
  hello["say"]("hello world");  
  int version=hello["version"].as<int>(); 
  auto person= hello["Person"]("me");
  std::cout<< person.call("intro") <<"\n";  
  return 0;
}
\end{lstlisting}

To help users import and use Svar modules in Python, a open source package is provided to load the shared libraries as python modules.
Developers can install the tool with pip and directly use the loaded module just like traditional python packages:

\begin{lstlisting}[language=python,caption={Example Python code to import and use Svar module.},label=code_demo]
import svar # import the loading tool
hello = svar.load("hello")

hello.say("hello world")
version=hello.version
person=hello.Person("me")
print(version,person.intro())
\end{lstlisting}

Unlike other binding tools where Python dependency is tightly coupled with module implementations, Svar modules do not rely on extra dependencies. The same Svar module can be imported by different Python versions (or even languages) at the same time, while developers have to compile a shared library for each version using other binding tools.

We also provides a bridge tool for Node.JS based on the NAPI of V8 JavaScript engine so that Svar objects can be imported as internal modules seamlessly. 

\begin{lstlisting}[language=JavaScript,caption={Example Javascript code to import and use Svar module.},label=code_demo]
hello=require("./svar")("hello")

hello.say("hello world")
console.log(hello.version)

person=new hello.Person("me")
console.log(person.intro())
\end{lstlisting}

By using Svar, users do not need change anything to call the same built library in different languages, and the bridge tool is also general for all Svar plugin modules.
More bridge tools for other programming languages can be implemented though their interface to C/C++.

\begin{figure*}[tb]
	\centering
	\includegraphics[width=0.99\textwidth]{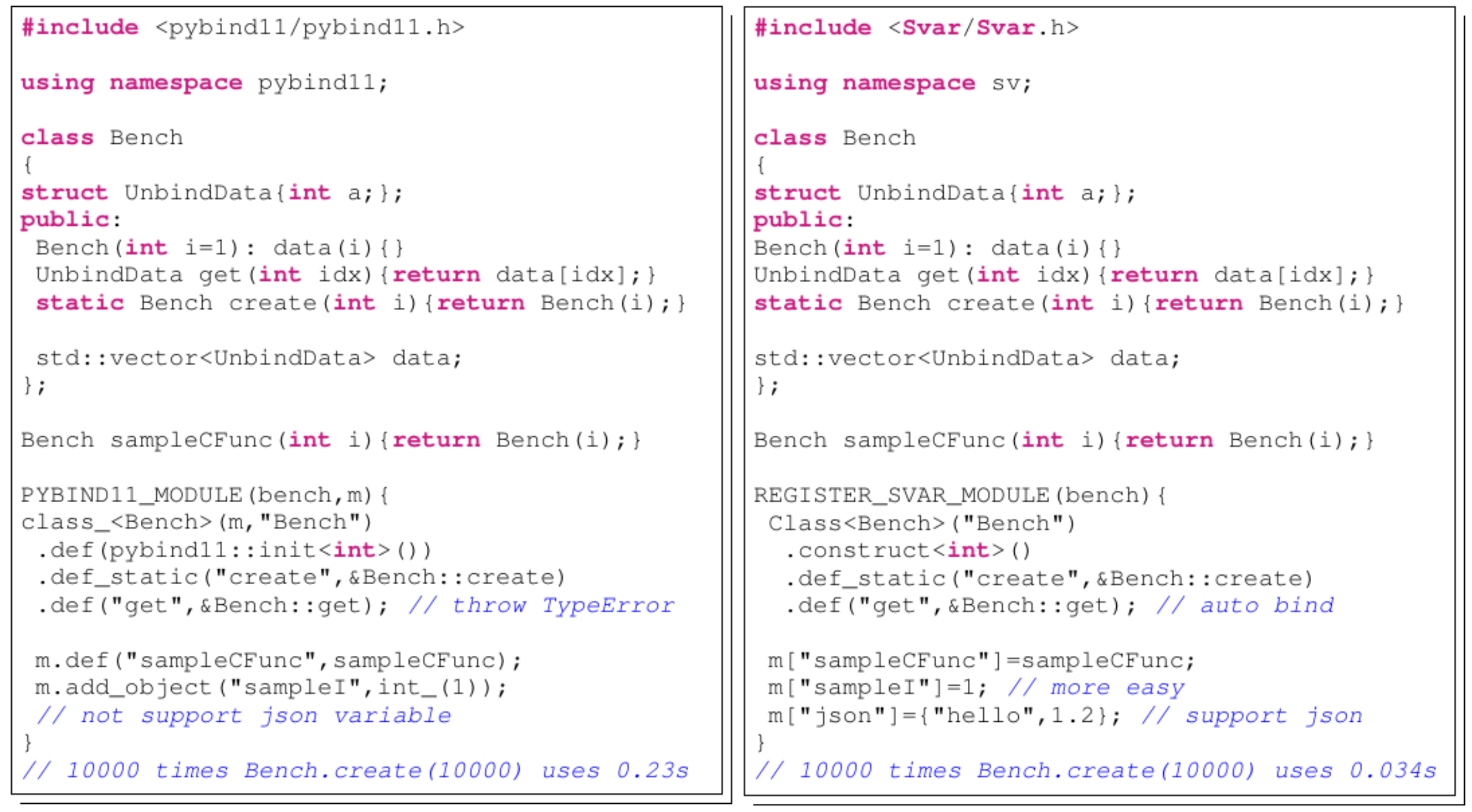}
	\caption{Comparison between Svar and pybind11 for binding class, function and variables.
		It shows that the usage of Svar is as easy as pybind11, while the function and variable binding can be even more direct.
		Benefit from the efficient design, Svar can be over 7 times faster than pybind11 in some circumstances,
		and Svar auto bind all classes used while pybind11 unable to convert unknown instances to Python type.
	}
	\label{fig:compare}
\end{figure*}

\subsection{Bridge for Languages} 

Svar not only help exporting C++ libraries to other languages, but also able to bridge those languages inversely.
Here below is a sample code to import and use Python in C++, as Svar is used to decouple the low level of languages, programmers are able to use a library implemented by other scripting languages as usual Svar modules.

\begin{lstlisting}[caption={Sample code to import and use Python modules in C++.},label=code_demo]
#include <Svar.h>

int main(){
  auto python = svar.import("svarpy");
  auto os = python["import]("os");
  std::cout<<"Pid is:"<<os["getpid"]();
  return 0;
}
\end{lstlisting}

As we are able to export and import libraries implemented by different languages, Svar can also be used as a bridge to import and call modules across languages. Here below is a sample code to import and use Python module in Node.JS:

\begin{lstlisting}[language=JavaScript,caption={Sample code to import and use Python module in Node.JS.},label=code_demo]
python=require("./svar")("svarpy")
os=python.import("os")
print("Pid is", os.getpid())
\end{lstlisting}

The above code firstly load a Python module as Svar object, and then translate it to the Javascript module, which can be used as usual module seamlessly. Svar provides a tiny but high performance bridge between programming languages.

Every programming language has its eco-system, it would be nice if we can fuse them to a larger one.
Some languages lack plenty third-party libraries and becomes very difficult for developers to develop some applications.
Once the interfaces to Svar modules implemented, it could be much easier as rich choices are provided from Svar by importing libraries of other languages.

\section{Experiments} \label{sec_exp}

Svar is tiny but elegant, which not only provides a general interface for multiple programming languages, the dynamic features also helps C++ for more easier implementation of several applications.
In this section, we firstly compare Svar with other existing popular language binding tools pybind11, nbind in terms of experience and efficiency.
And then, several applications is introduced to demonstrate the benefits using Svar in C++ implementations.
All experiments is conducted to shows that Svar is easy, efficient and ready-for-applications.

\subsection{Comparison to Other Binding Tools}

There are many binding tools like SWIG \cite{beazley1996swig}, pybind11 \cite{jakob2017pybind11}, nbind \cite{charto2017nbind} developed to help the interaction between scripting languages and C++.
Among them pybind11 is the most popular open-source library that is now used by the well-known deep learning projects pytorch \footnote{https://pytorch.org/} and tensorflow \footnote{https://www.tensorflow.org/} (transfered from SWIG in 2019).

We provide a little comparison between Svar and pybind11 for binding class, function and variables in Fig. \ref{fig:compare}.
The code comparison shows that the usage of Svar is almost as the same of pybind11 and even more straightforward.
Moreover, comparing to pybind11, Svar has the following advantages:

\begin{enumerate}	
	\item \textbf{Dependency free and ready for multiple programming languages.} Pybind11 directly use the dynamism of Python and C++ implemented modules have to compile rely on the Python dependency. This also means the modules can not be used across different Python versions. Svar however, does not brings any extra dependency to the C++ implementation and modules can be used across different Python versions and programming languages. This brings great convenience for C++ applications where Python is only used for module testing, as we do not need to separately provide a wrapper library.	
	
	\item \textbf{Support Json style complex data structure.} Pybind11 supports lots of containers which are existing in STL. However, the elements has to keep all the same and it is hard to represent complex data structures where different containers are nested with a varias types. Svar is able to represent complex data structure in a JSON style easily. This can significantly simplify the API design in some circumstances such as parameter settings.
	
	\item \textbf{Auto bind everything used and is even more easier to use.} As illustrated in Fig. \ref{fig:compare}, pybind11 has to bind every used class explicitly, otherwise an exception will be raised. Svar instead auto binds everthing which are been hold, so that they can be further used as arguments.
	And exporting variables and functions can be even more easier in Svar as no paticular binding function or bridge classes are required.	
	
	\item \textbf{More light-weighted and efficient.} Although Svar does not use the dynamism of Python directly, the experiments shows it even brings higher efficiency (7 times faster in the demo case). Benifits from the light-weight design, the type conversion and function calling is even more direct and simple. 
	The dynamism is already done in Svar and the Python object directly hold a Svar pointer. 	
\end{enumerate}

Svar has potential to unify different binding tools since it is more lightweight, dependency free, compatible friendly and easy to use.
Comparing to existing binding tools, Svar not only able to provide an unified binding functional between scripting lanuages and C++ without dependency, but also forms a general bridge for multiple languages, which enables importing and calling across lanuages. 
Moreover, for C++ only developers, Svar also enable dynamically library importing which simplified interface design, compile and release.

\subsection{Object Serialization}

\begin{figure*}[tb]
	\centering
	\includegraphics[width=1.0\textwidth]{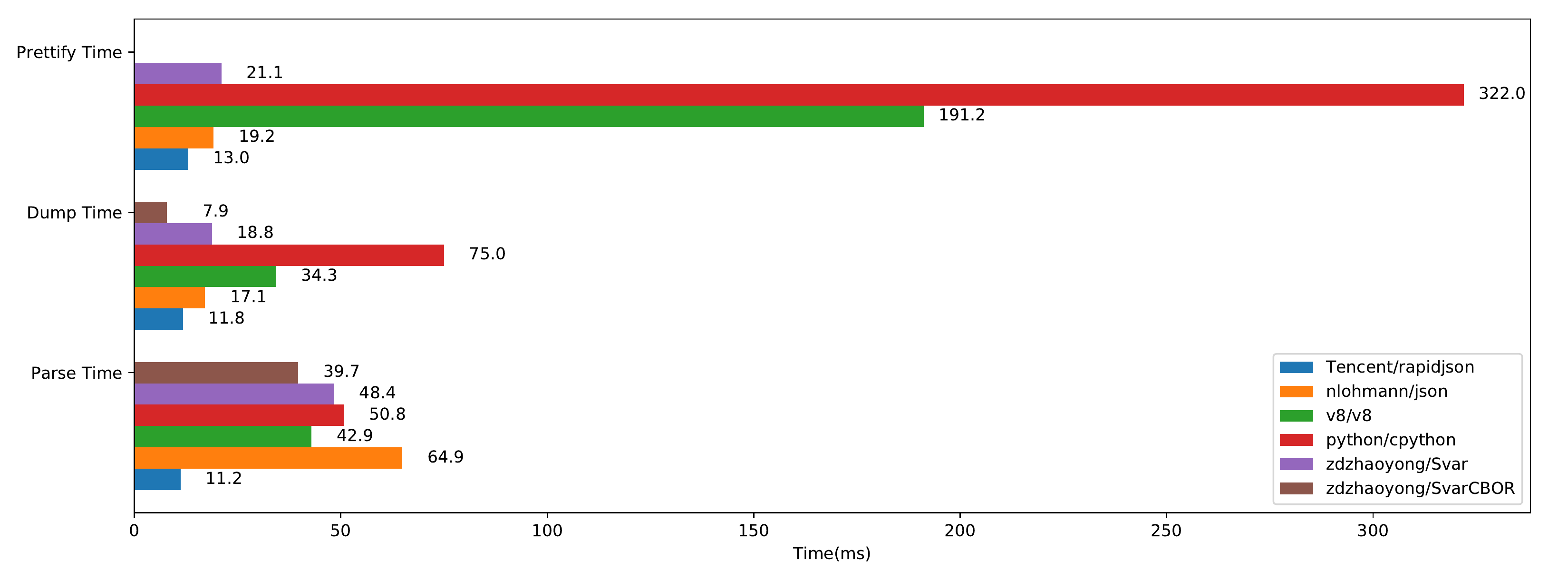}
	\caption{Performance comparison of different open-source C/C++ libraries with JSON parsing and generation capabilities.
		The total duration of serialize 3 JSONs on the public benchmark are illustrated. Lower is better.		
		Comparison between Svar and pybind11 for binding class, function and variables.
		Although Svar is not particular designed for serialization speed, the performance is comparable to state-of-the-art engines.
		By using binary JSON format CBOR, Svar achieves best serialize performance and can further support memory buffers.
	}
	\label{fig:jsontime}
\end{figure*}

Expect for the unified interface functional, the dynamism of Svar also brings more design patten in C++.
As we always need to save running data and results to file system or transfer them across computers, data serialization is one of the most basic functional in programming.
Protobuf \footnote{https://github.com/protocolbuffers/protobuf} provides an efficient contract-first protocol for serializing structured data, however, messages must be specified in .proto files when the application is built, which is not suitable for unstructured payloads.
JSON instead, is the most popular unstructured data format for storing and exchanging data, which is both easy for humans to read-write and also easy for machines to parse and generate.
Svar builtin supports JSON parse and dump, and Table \ref{fig:compare} shows the efficiency comparison of different existing popular engines.
JSON files on public benchmark \footnote{https://github.com/miloyip/nativejson-benchmark} are used.

%
%
%
%
%
%
%
%
%
%
%
%
%

It should be noted that although Svar is not particular designed for serialization speed, the performance is comparable to other engines, and Svar is single-header only with less than 5k lines code.
Moreover, Svar further supports memory buffers and serialize unstructured data to binary JSON formats such as CBOR (\url{https://cbor.io/}).
The serialization of memory buffers with unstructured data is important for contents like images and matrices, which is more efficient than using string to represent buffers with Base64 or Hex.
Table \ref{fig:compare} illustrates that the serialization of CBOR is faster than JSON, and further acceleration can be achieved by using buffers to hold structured data.

Except for the basic data structures, user defined objects are also supported for elegant serialization and restore.
Developers can either re-implement the caster template or define the buffer constructor and converter functions.

\begin{lstlisting}[caption={Sample code to dump and load user defined structs with CBOR.},label=code_demo]
#include <Svar.h>
using namespace sv;

Svar cbor=svar.import("svar_cbor");

struct Point2d{ double x,y;};

int main(){
 Class<Point2d>("Point2d")
  .def("__init__",[](SvarBuffer buf){
   Point2d ret;
   memorycpy(&ret.x,buf.ptr(),sizeof(ret));
   return ret; 
  })
  .def("__buffer__",[](Point2d& pt){
   return SvarBuffer((void*)&pt,sizeof(pt));
  });
  
 Svar sum=[](Point2d l,Point2d r)->Point2d{
  return {l.x+r.x,l.y+r.y};
 };
  
 Svar pts={Point2d({1,2}),Point2d({3,4})};
 auto buffer=cbor["dump"](pts); // serialize
  
 Svar r=cbor["load"](buffer); // restore
 Point2d result=sum(r[0],r[1]).as<Point2d>();
  
 assert(result.x==4);
 assert(result.y==6);
 return 0;
}
\end{lstlisting}

The above sample shows that Svar seamlessly combines structured objects with unstructured JSON format, and user defined data structure are serialized to and from buffer. 
Benefit from the dynamism of Svar, function calling can be performed directly using the restored objects with automatic type casting.

\subsection{HTTP RESTful API Module}

HTTP (Hypertext Transfer Protocol) is the foundation of data communication for the World Wide Web, where hypertext documents include hyperlinks to other resources that the user can easily access, for example by a mouse click or by tapping the screen in a web browser.
By using Flask \footnote{https://github.com/pallets/flask} web framework, Python developers can easily write a small server.
However, if a developer knows nothing about HTTP, writing a HTTP server or client is not so easy as it does in scripting languages.

An tiny open-source Svar module, \textit{svar\_http} \footnote{https://github.com/zdzhaoyong/svar\_http} brings new opportunity to provide and use HTTP service easily.
Developers does not need to know any detail about HTTP, and Svar modules can be directly exposed as network API.
Here below is a tiny demo to expose API through \textit{svar\_http}:

\begin{lstlisting}[caption={Sample code to expose API through HTTP Server.},label=code_demo]
#include <Svar.h>
using namespace sv;

Svar http=svar.import("svar_http");

int sum(int a,int b){return a+b;}

int main(){
 // direct serve a module
 helloM=svar.import("hello")
 helloS=http["Server"](helloM,"0.0.0.0:1233")

 // local defined service
 Svar api;
 api["print"]=[](Svar json){return json;} 
 api.def("sum",sum,"a"_a,"b"_b=0);
 
 server= http["Server"](api);
 server.call("listen","0.0.0.0:1234");
 return 0;
}
\end{lstlisting}

The above code demonstrates that a Svar module can be directly served through HTTP, and writing a server is very similar to exporting a Svar module, which is easy to understand and developers do not need to known any detail about the implementation of HTTP. 

\begin{lstlisting}[caption={Sample client code to interact with HTTP server.},label=code_demo]
#include <Svar.h>
using namespace std;

auto http=svar.import("svar_http");

int main(){
 std::string url="http://0.0.0.0:1234/";
 assert(http["get"](url+"sum?a=1&b=2")==3);
 cout<<http["get"](url+"print?args=hello");
 
 Svar body={{"a",1},{"b",2}};
 assert(http["post"](url+"sum",body)==3);
 cout<<http["post"](url+"print",body);
 return 0;
}
\end{lstlisting}

Interact with HTTP server is also very easy through the Svar interface.
Benefit from the powerful dynamism of Svar, the network API supports keyword parameters and arguments are automatically casted to the target type.

\subsection{Topic based Data Distrubution}

ROS \cite{quigley2009ros} is an open-source robot operating system which is popular used in the robotic community for both commercial or noncommercial usages. It provides a structured communications layer above
the host operating systems of a heterogenous compute cluster.
However, the design of ROS is too heavy and users have to define the strictly typed data structure of messages for communication.
Based on Svar, GSLAM \cite{gslamICCV2019} implemented a header-only intra-process communication utility class named Messenger, which is more light-weighted, easy-to-use and supports to publish and subscribe any class without
extra cost.
Moreover, data distribution can also be further bridged through network by integrate existing messaging protocols such as Kafka (\url{http://kafka.apache.org/}), MQTT (\url{https://mqtt.org/}) and NSQ (\url{https://nsq.io/}).

\section{Conclusion} \label{sec_con}

This paper presents a tiny programming languages interaction framework, which not only brings an unified binding tool for C++, but also forms a high performance bridge for multiple programming languages.
Moreover, the dynamism brings better design pattern for various applications and enables importing modules dynamically in C++.
The light-weight design makes it a tiny core to decouple and link library modules or even programming languages.
We make it open-source on github and free for both commercial or noncommercial usages, and more Svar modules will be developed with supporting to more languages in the future.



\bibliographystyle{IEEEtran} 
\bibliography{refs}

\end{document}